# Suspicious-Taint-Based Access Control for Protecting OS from Network Attacks

Zhiyong Shan

Department of Computer Science, Renmin University of China

**Abstract.** Today, security threats to operating systems largely come from network. Traditional discretionary access control mechanism alone can hardly defeat them. Although traditional mandatory access control models can effectively protect the security of OS, they have problems of being incompatible with application software and complex in administration. In this paper, we propose a new model, Suspicious-Taint-Based Access Control (STBAC) model, for defeating network attacks while being compatible, simple and maintaining good system performance. STBAC regards the processes using Non-Trustable-Communications as the starting points of suspicious taint, traces the activities of the suspiciously tainted processes by taint rules, and forbids the suspiciously tainted processes to illegally access vital resources by protection rules. Even in the cases when some privileged processes are subverted, STBAC can still protect vital resources from being compromised by the intruder. We implemented the model in the Linux kernel and evaluated it through experiments. The evaluation showed that STBAC could protect vital resources effectively without significant impact on compatibility and performance.

## 1 Introduction

With the rapid development and increasing use of network, threats to operating systems mostly come from network, such as buffer overflows, viruses, worms, Trojans, DOS, and etc. On the other hand, as computers, especially PCs, become cheaper and easier to use, people prefer to use computers exclusively and share information through network. Though on a few occasions a user may permit someone else who is fully trusted to log in his/her computer from local, most of the time users share information via network. Therefore nowadays the threat to modern OSs does not come from local, but more from remote.

Traditional DAC (Discretionary Access Control) in OS alone cannot defeat network attacks well. Traditional MAC (Mandatory Access Control) is effective in maintaining security, but it has problems of being incompatible with application software and complex in administration [1][2][3]. From 2000 to 2003, we developed a secure OS, which implemented BLP [5], BIBA [6], RBAC [7] and ACL. However, we found the same problems with the secure OS. Thus, the STBAC model is proposed with the following goals in mind.

- Protecting vital resources: Even if some privileged processes are subverted, STBAC can still protect vital resources from being compromised. Vital resources in OS include important binary files, configuration files, directories, processes, user data, system privileges and other limited resources, such as CPU time, disk space, memory, network bandwidth and important kernel data structures. Since they are the valuable user data and foundation for OS to provide services, they usually become the final target of an intrusion. Even if an intruder gets the root identity, we can say that the intrusion has failed if the intruder cannot compromise the vital resources.
- Compatibility: STBAC-enhanced OS is compatible with existing application software.
- Simplicity: STBAC is easy to understand and administer.
- High performance: STBAC is implementable with high performance.

The STBAC model regards the processes using Non-Trustable-Communications as the starting points of suspicious taint, traces the activities of the suspiciously tainted processes by taint rules, and forbids the suspiciously tainted processes to illegally access vital resources by protection rules.

We implemented the STBAC model in the Linux kernel, and evaluated its capability of protecting vital resources, compatibility and system performance through experiments.

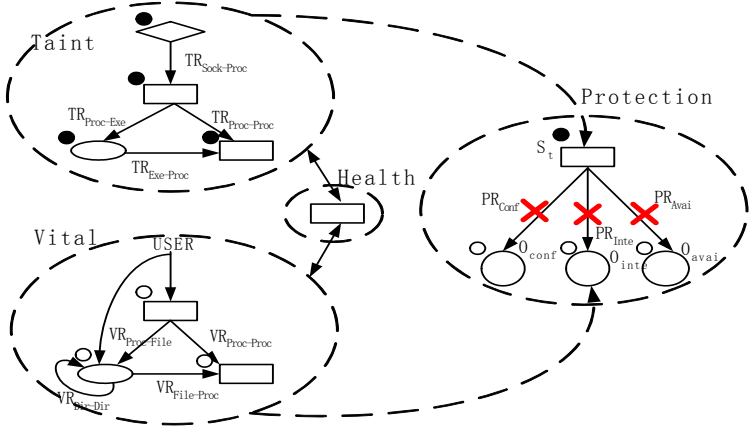

**Fig.1.** STBAC model

The paper is organized as follows. We first describe the STBAC model and its four parts in Section 2. In Section 3, the protection capability, compatibility and simplicity of STBAC are analyzed. Section 4 presents the implementation details of the STBAC model in the Linux kernel. The evaluation results are shown in Section 5. In Section 6, STBAC is compared with related works. Finally, we draw the conclusion in Section 7.

## 1. Model Description

The STBAC model consists of four parts: Taint, Health, Vital and Protection, as shown in Figure 1, where each part is enclosed in a dashed circle. The rectangles indicate processes; the ellipses indicate files or directories; the diamonds indicate sockets; and the balls indicate any entities in OS, such as files, directories, sockets and processes.

The Taint part, probably controlled by an intruder, consists of suspiciously tainted subjects ($S_t$), suspiciously tainted objects ($O_t$) and taint rules (TR). TR is categorized into $TR_{sock-proc}$, $TR_{proc-proc}$, $TR_{proc-exe}$ and $TR_{exe-proc}$, and any $S_t$ or $O_t$ in Figure 1 has a solid dot on its upper left. The Vital part represents the vital resources that should be protected properly. It consists of vital objects that include $O_{conf}$, $O_{inte}$ and $O_{avai}$, and vital rules (VR) that include $VR_{proc-proc}$, $VR_{dir-dir}$, $VR_{proc-file}$ and $VR_{file-proc}$. Any vital object in Figure 1 has a hollow dot on its upper left. The Protection part consists of three protection rules (PR): $PR_{conf}$, $PR_{inte}$ and $PR_{avai}$, which forbids $S_t$ to illegally access vital objects. The Health part consists of health objects ($O_h$) and health subjects ($S_h$) that are not tainted or labeled as vital ones. We elaborate on the four parts of STBAC in the following sections.

### 1.1 Taint

As the intruder probably controls the $S_t$ and $O_t$, STBAC labels them with suspiciously tainted flag ($F_t$), and traces $S_t$'s activities in OS kernel with taint rules.

*1.1.1 Taint Entities*

First of all, we define remote network communications with necessary security means as **Trustable-Communications**, e.g., the secure shell, and those without security means as **Non-Trustable-Communications**. Security means include authentication, confidentiality protection, integrity protection, and etc.

**Suspiciously Tainted Subject ($S_t$)** is a subject that may be controlled by an intruder and may act for intrusion purposes. $S_t$ is a process in general. For example, it can be a process using Non-Trustable-Communications, or a process of an executable file created by an intruder, or a process of an executable file downloaded from network, or the descendant process of the above processes. It can also be a process that communicates with the above processes, or a descendant of such a process.

**Suspiciously Tainted Object ($O_t$)** is an object that is created or modified by an intruder, and may aid in the intrusion. Generally, $O_t$ means either the executable file created and modified by $S_t$, or the process created and accessed by $S_t$, or the file and directory accessed by $S_t$.

Both $S_t$ and $O_t$ are labeled with **Suspiciously Tainted Flag ($F_t$)**.

**Table 1** Information flows

| Information flow | Operation |
|---|---|
| Process→process | Fork, signal, IPC |
| Process→file | Create, write |
| Process→dir | Create, write |
| Process→socket | Create, write |
| File→process | Read, execute |
| Dir→process | Read, execute |
| Socket→process | Read |

*1.1.2 Taint Rules*

Information flows between subjects and objects in OS are significant to OS security research [8] [9]. There have been several studies on the method of backtracking intrusions in kernel based on the dependency graph that depicts information flows [10~13]. With the dependency graph, the administrator can easily trace out all processes and files related to intrusion. STBAC adopts a similar approach to construct taint rules. The information flows that are possible to spread taint are depicted in Table 1.

If we build taint rules based on Table 1, the number of tainted processes, files and directories can be very large. The main reason is that $S_t$ will taint a vast number of $S_t$s and $O_t$s during its frequent file and directory operations. This may be exploited by the intruder to generate false dependencies [10], and eventually to aggravate system workload heavily.

Although there are a lot of file and directory operations, most of them cannot spread taint. File and directory operations that can spread taint fall into two types: 1) creating, writing and executing executable files; 2) reading and writing files that may influence process actions. These files are important configuration files or data files. For the second type of operations, we can avoid spreading taint by setting important configuration or data files as integrity protected ones. Thus we can forbid $S_t$ to write these files. So, we only need to treat the first type of operations as taint rule instead of the entire set of file and directory operations shown in Table 1. Therefore, we build taint rules as follows:

**Socket to Process Taint Rule** ($TR_{Sock\text{-}Proc} : Socket \xrightarrow{F_t} \Pr ocess$) depicts that the process using Non-Trustable-Communications may be breached or launched by the intruder. Thus it should be labeled with $F_t$. In contrast, the process using Trustable-Communications should not be labeled with $F_t$.

**Process to Process Taint Rule** ($TR_{Proc\text{-}Proc} : \Pr ocess \xrightarrow{F_t} \Pr ocess$) depicts that the process created by $S_t$ or received communication message from $S_t$ should be labeled with $F_t$. No doubt, the process created by $S_t$ is dangerous, so it is regarded as $S_t$. By means of pipe, local socket, shared memory or message queue, $S_t$ may control other process to serve for intrusion, thus the controlled process is also regarded as $S_t$.

**Process to Exe-file Taint Rule** ($TR_{Proc\text{-}Exe} : \Pr ocess \xrightarrow{F_t} Executable\ File$) depicts that the executable file created or modified by $S_t$ should be labeled with $F_t$. Executable files created by $S_t$ may be hostile programs, such as programs downloaded from network. On many occasions, modifying or over-writing existing executable files is a way to leave backdoor, for example, using specially modified "ls", "ps" and "netstat" to over-write existing command files.

**Exe-file to Process Taint Rule** ($TR_{Exe\text{-}Proc} : Executable\ File \xrightarrow{F_t} \Pr ocess$) depicts that the process that has executed or loaded $O_t$ should be labeled with $F_t$. Suspiciously tainted command files, library files, or other executable files could be intrusion tools, so the process derived from them is dangerous.

## 1.2 Vital

The Vital part is the target for STBAC to protect, which consists of vital objects and vital rules. Vital objects include all kinds of vital resources, such as important user data, important system files or directories, limited system resources, and etc. Vital rules define the conditions to spread vital flags that are used to label vital objects.

*1.2.1 Vital Entities*

According to the information protection targets proposed in ITSEC [14], even if OS is subverted, STBAC should still protect the following three types of objects:

**Confidentiality Object ($O_{conf}$)** is an object containing information that should be protected confidentially even if the system is breached. Generally, $O_{conf}$ means a file or directory containing sensitive information. For example, "/etc/passwd" and "/etc/shadow" are typical $O_{conf}$s. $O_{conf}$ is labeled with **Confidentiality Flag ($F_{conf}$)**.

**Integrity Object ($O_{inte}$)** is an object whose integrity should be protected even if the system is breached. Generally $O_{inte}$ means binary files, important configuration files, important user data files and directories containing these files. $O_{inte}$ is labeled with **Integrity Flag ($F_{inte}$)**.

**Availability Object ($O_{avai}$)** is the limited resource that is necessary to run processes. Even if the system is breached, OS should guarantee that $S_t$ do not block other vital or health processes getting $O_{avai}$. $O_{avai}$ includes CPU, memory, disk space, network bandwidth and important kernel structures. $O_{avai}$ is labeled with **Availability Flag ($F_{avai}$)**.

In order to perfect the confidentiality protection function of STBAC, we further introduce two definitions.

**Leak Object ($O_{leak}$)** is an executable file from which a process derived may leak secrecy while writing files after reading an $O_{conf}$. Typical examples are "cp", "mcopy", "dd", "passwd", and etc.

**Leak Subject ($S_{leak}$)** is a process derived from $O_{leak}$ that may leak secrecy while writing files after reading an $O_{conf}$.

Both $O_{leak}$ and $S_{leak}$ are labeled with **Leak Flag ($F_{leak}$)**.

*1.2.2 Vital Rules*

As presented above, STBAC identifies $O_{conf}$, $O_{inte}$, $O_{avai}$, $O_{leak}$ and $S_{leak}$ by vital flags of $F_{conf}$, $F_{inte}$, $F_{avai}$ and $F_{leak}$ respectively. Before running, OS configures vital flags by default or by the administrator; in running, vital flags should be spread automatically to avoid security vulnerability. Thus, four rules for spreading vital flags are designed as follows:

**Directory to Directory Vital Rule** ($VR_{Dir\text{-}Dir} : Directory \xrightarrow{F_{conf}, F_{inte}} Directory$) depicts that the new directory or file inherits $F_{conf}$ and $F_{inte}$ from the parent directory at the creation time.

**Process to Process Vital Rule** ($VR_{Proc\text{-}Proc} : \Pr ocess \xrightarrow{F_{conf}, F_{leak}} \Pr ocess$) depicts that the new process inherits $F_{conf}$ and $F_{leak}$ from the parent process at the creation time.

**Process to File Vital Rule** ($VR_{Proc\text{-}File} : \Pr ocess \xrightarrow{F_{conf}} File$) depicts that any file should inherit $F_{conf}$ when it is created or modified by a process that has been labeled with $F_{conf}$ and $F_{leak}$ simultaneously.

**File to Process Vital Rule** ($VR_{File\text{-}Proc} : File \xrightarrow{F_{conf}, F_{leak}} \Pr ocess$) depicts that any $S_h$ should clean old $F_{conf}$ and $F_{leak}$ flags when executing a file, and then should inherit $F_{leak}$ from the executable file. In addition, any $S_h$ should possess $F_{conf}$ after reading $O_{conf}$.

## 1.3 Health

The Health part consists of health objects ($O_h$) and health subjects ($S_h$). A **Health Subject ($S_h$)** is a process that has not been tainted or labeled as vital. A **Health Object ($O_h$)** is an object that has not been tainted or labeled as vital. The Health can access the Taint and the Vital, and vice versa.

## 1.4 Protection

Corresponding to the three security protection targets, confidentiality, integrity and availability, STBAC sets up three protection rules，which constitute the Protection part.

**Confidentiality Protection Rule** ($PR_{Conf} : S_t \xrightarrow{R} | O_{conf}$) forbids $S_t$ to read $O_{conf}$, i.e. it forbids suspiciously tainted subjects to read sensitive files, to read or search sensitive directories, and to execute some privileged operations to destroy confidentiality, such as the "ptrace" system call.

**Integrity Protection Rule** ($PR_{Inte} : S_t \xrightarrow{W} | O_{inte}$) forbids $S_t$ to write $O_{inte}$, i.e. it forbids suspiciously tainted subjects to modify, create, delete and rename a protected file or directory, and to execute some privileged operations to destroy integrity, such as the "create_module" and "setuid" system calls.

**Availability Protection Rule** ($PR_{Avai} : O_{avai} \xrightarrow[R_{St} > HWM_{St} \text{ or } R_{SYS} > HWM_{SYS}]{A} | S_t$) forbids an $O_{avai}$-allocating operation if the operation could result in that the amount of allocated $O_{avai}$ exceeds one of the two High Water Markers (HWM). One HWM is named $HWM_{St}$, which represents maximum amount of $O_{avai}$ permitted for $S_t$ to get. The other HWM, named $HWM_{SYS}$, represents maximum percentage of allocated $O_{avai}$ in the whole system. ($R_{St}$ denotes the amount of $O_{avai}$ allocated to $S_t$; and $R_{SYS}$ denotes the percentage of allocated $O_{avai}$ in the whole system.)

If treating STBAC and its members as sets, we can formally represent the STBAC model as follows:

$$STBAC = \{Taint, Health, Vital, Protection\}$$

$$Taint = \{S_t, O_t, TR\}$$

$$TR = \{TR_{Sock\text{-}Proc}, TR_{Proc\text{-}Proc}, TR_{Proc\text{-}Exe}, TR_{Exe\text{-}Proc}\}$$

$$TR_{Sock\text{-}Proc}: Socket \xrightarrow{F_t} Process, \quad TR_{Proc\text{-}Proc}: Process \xrightarrow{F_t} Process$$

$$TR_{Proc\text{-}Exe}: Process \xrightarrow{F_t} Executable\ File, \quad TR_{Exe\text{-}Proc}: Executable\ File \xrightarrow{F_t} Process$$

$$Health = \{S_h, O_h\}$$

$$Vital = \{O_{conf}, O_{inte}, O_{avai}, O_{leak}, S_{leak}, VR\}$$

$$VR = \{VR_{Dir\text{-}Dir}, VR_{Proc\text{-}Proc}, VR_{Proc\text{-}File}, VR_{File\text{-}Proc}\}$$

$$VR_{Dir\text{-}Dir}: Directory \xrightarrow{F_{conf}, F_{inte}} Directory, \quad VR_{Proc\text{-}Proc}: Process \xrightarrow{F_{conf}, F_{leak}} Process$$

$$VR_{Proc\text{-}File}: Process \xrightarrow{F_{conf}} File, \quad VR_{File\text{-}Proc}: File \xrightarrow{F_{conf}, F_{leak}} Process$$

$$Protection = \{PR\} = \{PR_{Conf}, PR_{Inte}, PR_{Avai}\}$$

$$PR_{Conf}: S_t \xrightarrow{R} \Big| O_{conf}, \quad PR_{Inte}: S_t \xrightarrow{W} \Big| O_{inte}, \quad PR_{Avai}: O_{avai} \xrightarrow[R_{St} > HWM_{St}\ or\ R_{SYS} > HWM_{SYS}]{A} \Big| S_t$$

## 2. Model Analysis
### 2.1 Protection Analysis

*2.1.1 Confidentiality*

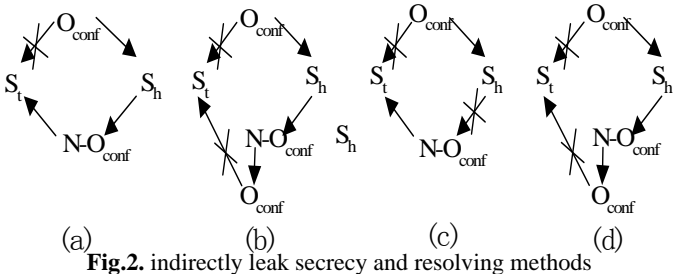

**Fig.2.** indirectly leak secrecy and resolving methods

$PR_{conf}$ of STBAC prevents $S_t$ from reading $O_{conf}$ for preserving confidentiality. However, as $PR_{conf}$ does not restrict $S_h$, there possibly exists an information flow that indirectly leaks secrecy, as shown in Figure 2(a). Although $S_t$ cannot get secrecy information from $O_{conf}$ directly, $S_t$ can get it indirectly by the path of $O_{conf} \rightarrow S_h \rightarrow N\text{-}O_{conf} \rightarrow S_t$. ($N\text{-}O_{conf}$ denotes an object that has no secrecy, such as $O_h$, $O_t$, and other objects).

This indirect leakage requires the participation of $S_h$, and it will not occur without $S_h$. In other words, it is difficult to leak secrecy only under remote user's attack and without local user's cooperation or misuse. The reason lies in that $S_t$ cannot control $S_h$ and shake off the tracing of taint rules simultaneously. Sometimes, $S_t$ wants to control a $S_h$. By the operations of creating processes or executing files, or by IPC communications, $S_t$ may succeed in controlling a $S_h$. However, according to the taint rules, these operations are bound to taint the $S_h$.

In a system that users have a sense of security, the indirect leakage of sensitive information won't happen. Nevertheless, we present three ways to prevent leaking sensitive information via user's carelessness.

- Relay-spread. After a $S_h$ reads $O_{conf}$, $F_{conf}$ is spread to $S_h$. While $S_h$ further writes to an $N\text{-}O_{conf}$ file, $F_{conf}$ will be relay-spread to the file. Thus, by $PR_{conf}$, $S_t$ cannot read information from the file any more. Figure 2(b) shows this strategy.
- Forbid-writing. After a $S_h$ reads $O_{conf}$, $F_{conf}$ is spread to $S_h$. If $S_h$ wants to further write to an $N\text{-}O_{conf}$ file, we forbid it. Figure 2(c) shows this strategy.
- Selective-spread. After a $S_h$ reads $O_{conf}$, $F_{conf}$ is spread to $S_h$. When $S_h$ further writes to an $N\text{-}O_{conf}$ file, $F_{conf}$ will be spread to the file only if the writing may cause secrecy leakage. Thus, by $PR_{conf}$, $S_t$ cannot read the file any more. Figure 2(d) shows this strategy.

From these figures, we can easily find that all three strategies can effectively cut off leaking paths. For relay-spread, there can be the problem of false leakage of sensitive information, i.e. after reading $O_{conf}$, the writing operation may not necessarily write sensitive information to $N\text{-}O_{conf}$. False leakage of sensitive information will bring rapidly many $O_{conf}$s into the system, however. As $S_t$ cannot read $O_{conf}$s, $S_t$'s behavior will be restricted so that $S_t$ cannot run normally.

For forbid-writing, as it forbids writing from $S_h$ to $N\text{-}O_{conf}$, we can rationally consider that $S_h$ with $F_{conf}$ has higher sensitive level than $N\text{-}O_{conf}$. This strategy is similar to the usual BLP model enforcement [15][16]. Certainly, forbid-writing can avoid the problem of false leakage of sensitive information, but to some extent this may also restrict $S_h$'s behavior so that $S_h$ cannot run normally.

For selective-spread, it always permits writing to $N\text{-}O_{conf}$, no matter whether it causes spreading $F_{conf}$ to $N\text{-}O_{conf}$ or not, hence it does not disturb $S_h$'s running. At the same time, it still can prevent leaking secrecy via spreading $F_{conf}$ to $N\text{-}O_{conf}$ when the writing may cause secrecy leakage. So the selective-spread approach is chosen in our model.

The selective-spread strategy can be implemented using $F_{leak}$, $VR_{proc\text{-}file}$ and $VR_{file\text{-}proc}$ that are described in Section 2. For instance, copying passwd file from /etc to /tmp can cause leaking secrecy. After executing the cp command file that is labeled with $F_{leak}$ in advance, the process will inherit $F_{leak}$ from the cp file according to $VR_{file\text{-}proc}$; when the process reads the passwd file labeled with $F_{conf}$, $F_{conf}$ will be spread to the process by $VR_{file\text{-}proc}$; when the process subsequently creates a new file /tmp/passwd, $F_{conf}$ will be further spread to the new file, thus the secrecy information in the new file will undoubtedly be protected. However, using the same cp command won't spread $F_{conf}$ if user copies an $N\text{-}O_{conf}$ file to anywhere. Only if a process has both $F_{conf}$ and $F_{leak}$ can it spread $F_{conf}$.

In summary, STBAC can prevent directly and indirectly leaking secrecy based on $PR_{conf}$ and selective-spread mechanism, so it can protect confidentiality well.

*2.1.2 Integrity*

According to the $PR_{inte}$ and taint rules, STBAC can meet the three conditions of the "Low-Water Mark Policy for Objects" in Biba's model [6], which are:

(1) Any subject (S) can read any object (O), regardless of their integrity levels (i). After reading, there will be i(s) = min(i(s), i(o)).
(2) s∈S can write o∈O at any integrity level. After writing, there will be i(o) = min(i(s),i(o));
(3) s1∈S can execute s2∈S, only if i(s2)≤i(s1).

By $PR_{inte}$, we can reasonably confirm that $i(S_t)=i(O_t)<i(O_{inte})=i(S_h)=i(O_h)$. So the first condition can be satisfied by taint rules of $TR_{sock\text{-}proc}$, $TR_{proc\text{-}proc}$ and $TR_{exe\text{-}proc}$. The second condition can only be satisfied partially by the taint rule of $TR_{proc\text{-}exe}$, because, after a $S_t$ write to a $O_h$, STBAC will not downgrade $i(O_h)$ when the $O_h$ is not an executable file, this does not comply with the second condition. However, this will not be exploited by intruder for that $PR_{inte}$ can protect important executable files, configuration files and data files from being written by $S_t$. The third condition requires that low integrity level subject cannot execute high integrity level object, i.e. $S_t$ cannot get a $S_h$ if $i(S_t)<i(S_h)$. In detail, according to $TR_{proc\text{-}proc}$ and $TR_{exe\text{-}proc}$, though $S_t$ can get a process by operations of creating processes or executing files, the process gotten by $S_t$ surely is labeled with $F_t$. Thus the process gotten by $S_t$ is also a $S_t$, but not a $S_h$. So, the third condition is met.

In summary, STBAC basically satisfies the "Low-Water Mark Policy for Objects", so that it can prevent integrity damage caused by intruder.

*2.1.3 Availability*

Availability protection is to prevent illegal blocking of accessing data or services [14]. The action tampering availability has been concluded as DOS [17]. STBAC can protect availability in two manners:

First, from the perspective of OS, availability has to be built on the basis of integrity. Hence, STBAC protects firstly integrity of data, service configurations and system configurations by $PR_{inte}$.

Second, STBAC restricts allocation of resources by two HWMs, i.e. $HWM_{St}$ and $HWM_{SYS}$. $HWM_{St}$ restricts the amount of resources allocated to each $S_t$. $HWM_{SYS}$ forbids allocating new resources to $S_t$ when idle resources in the system is very few.

However, availability protection still requires that the system can tolerate or recover from internal errors, external attacks, and even physical failures. Due to the shortage of access control mechanism, these requirements have to be met by error-recovery [18] or self-healing [19] mechanisms.

## 2.2 Compatibility Analysis

STBAC does not influence the actions of local users and remote users using Trustable-Communications. It also does not affect most actions of $S_t$, because STBAC only forbids $S_t$ to illegally access vital resources, which are merely a small part of all the resources, and does not forbid $S_t$ to legally access vital resources, such as reading and executing $O_{inte}$.

Possible incompatibility can be caused by $PR_{conf}$ and $PR_{inte}$ since they restrict processes' actions. But they do not restrict the user who logs in by Trustable-Communications. This means that the administrator can still manage the computer and upgrade application software remotely by Trustable-Communications. $PR_{avai}$ only restricts the resource allocation, and it'll not restrict any normal action of a process if the two High Water Markers are configured properly.

On most occasions, reading $O_{conf}$ and modifying $O_{inte}$ through Non-Trustable-Communications mean intrusions or network worms, and these should be forbidden by $PR_{conf}$ and $PR_{inte}$. However, on special cases, we should permit processes using Non-Trustable-Communications to read $O_{conf}$ or modify $O_{inte}$, which we call Shared-$O_{conf}$ and Shared-$O_{inte}$ respectively. And they introduce incompatibility.

Shared-$O_{inte}$ is usually a system configuration file that has to be modified by a process using Non-Trustable-Communications. Furthermore, the process cannot change to use Trustable-Communications. So the amount of Shared-$O_{inte}$ is tiny. Shared-$O_{inte}$ can not be a binary file, application configuration file or the majority of system configuration files, because we can use Trustable-Communications such as SSL, TSL and SSH to upgrade the system, patch software and modify configurations remotely. Only exceptional system configuration files have to be modified through Non-Trustable-Communications. In Linux, Shared-$O_{inte}$ means /etc/resolve.conf, because dhcplient will write /etc/resolve.conf after receiving information from the remote DHCP server, whereas the communication between the DHCP client and server cannot use authentication or encryption.

Timothy Fraser successfully resolved a problem like Shared-$O_{inte}$ by setting trusted program [3]. Here we use a similar mechanism named Trustable-Communication-List, each entry of which consists of local program name, local IP and port, remote IP and port, network protocol and permitted time span. Information of remote communications that are needed when modifying Shared-$O_{inte}$ is put in the Trustable-Communication-List. Only when a remote communication, which is ready to be launched, matches an entry in the list will it be regarded as trustable. This mechanism can resolve Shared-$O_{inte}$ problem and assure security to some degree.

Shared-$O_{conf}$ are mainly the password files whose secrecy has to be shared by local processes and processes using Non-Trustable-Communications. Thus, the amount of Shared-$O_{conf}$ is tiny. In Linux, Shared-$O_{conf}$ includes /etc/passwd, /etc/shadow and /usr/local/apache/passwd/passwords.

A mechanism named Partial-Copy is designed to resolve the Shared-$O_{conf}$ problem. It generates a partial copy for each Shared-$O_{conf}$ to save part of the Shared-$O_{conf}$ content. The partial copy permits access by the processes using Non-Trustable-Communications. For example, we can build a partial copy of /etc/passwd to contain user information needed by process using Non-Trustable-Communications, but the information of privileged users and other important users stay in /etc/passwd and is forbidden to be accessed by Non-Trusted-Communication processes. In order to implement the Partial-Copy mechanism, the kernel should redirect the access target of Non-Trustable-Communication processes from Shared-$O_{conf}$ to the corresponding partial copy.

In summary, STBAC can get good compatibility because it only prevents $S_t$ from illegally accessing vital resources. Though we have incompatibility problems from Shared-$O_{conf}$ and Shared-$O_{inte}$, the amounts of these objects are tiny, and they can be resolved by the Trustable-Communication-List and Partial-Copy mechanisms.

## 2.3 Simplicity Analysis

Simplicity of STBAC derives from the fact that it is simple to administer and easy to understand. The main work for administering STBAC is to identify those files or directories that need to be protected and set vital flags. This is straightforward and easy to understand. As the system files and directories that need protection could be set vital flags automatically by the system, the user only needs to set his/her data files and directories. Taint flag can be generated and spread automatically by the kernel, and does not need any manual operations. Partial-Copy and Trustable-Communication-List may bring some additional work, but the work is limited because of the very small amount of Shared-$O_{conf}$ and Shared-$O_{inte}$.

## 3. Model Implementation

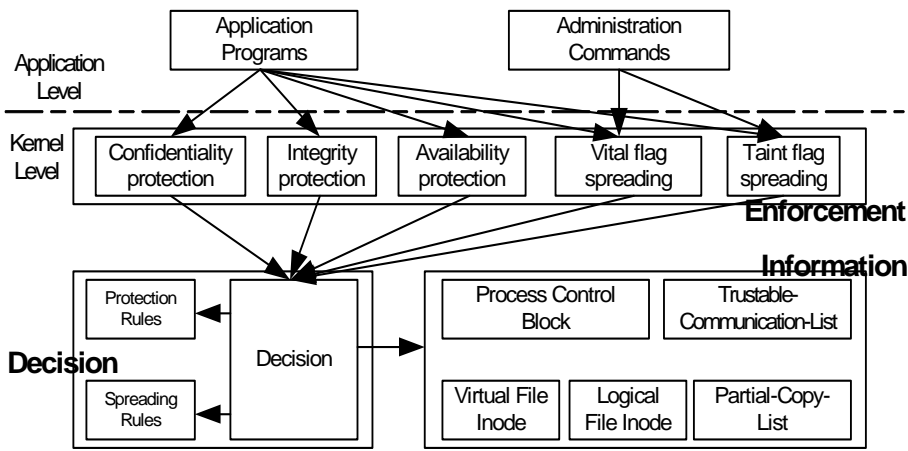

**Fig.3.** the enforcement of STBAC in Linux kernel

We have implemented a STBAC prototype in the Linux kernel 2.4.20 based on our former work. The general principle is to avoid significant reductions of simplicity, compatibility and performance of original Linux. Figure 3 shows the architecture. The total amount of the codes is less than five thousand lines, and most of them are located in kernel.

Similar to the methodology of M. Abrams et al. [23], we divide the implementation into three parts: enforcement, decision and information. Separating model enforcement from model decision has the advantage of conveniently modifying and adding model rules without change of most codes, as described in [24]. The information part is not independent of the kernel, but is founded on modifying existing kernel structure.

The enforcement part intercepts accesses at related system calls or important kernel functions, and issues requests to the decision part. For the protection requests, such as confidentiality protection requests, integrity protection requests or availability protection requests, the enforcement part permits or denies the access according to the result returned by the decision part. For the spread requests, such as taint spread requests and vital flag spread requests, the enforcement part does nothing after posting the requests, and the decision part directly modifies data structures of the information part. Table 2 describes the modified system calls and kernel functions, and the corresponding model rules.

**Table 2** STBAC rules and system calls

| Model rules | | Functions |
|---|---|---|
| Taint Rules | $TR_{sock-proc}$ | sys_socket |
| | $TR_{proc-proc}$ | sys_fork,sys_vfork,sys_clone,sys_pipe,sys_map,sys_shmat,sys_msgrcv,sys_mkfifo,sys_mknod |
| | $TR_{proc-exe}$ | sys_open,sys_create,sys_chmod,sys_fchmod |
| | $TR_{exe-proc}$ | sys_execve,sys_mmap |
| Vital rules | $VR_{dir-dir}$ | sys_open,sys_create,sys_mkdir,sys_mknod |
| | $VR_{proc-proc}$ | sys_fork,sys_vfork,sys_clone |
| | $VR_{file-proc}$ | sys_execve |
| | $VR_{proc-file}$ | Sys_open, sys_create |
| Protection rules | $PR_{conf}$ | sys_open,sys_ptrace,sys_get_stbac_attr |
| | $PR_{inte}$ | sys_open,sys_truncate,sys_ftruncate,sys_chmod,sys_fchmod,sys_chown,sys_fchown,sys_lchown,sys_rmdir,sys_rename,sys_unlink,sys_mount,sys_umount,sys_setrlimit,sys_reboot,sys_swapoff,sys_create_module,sys_delete_module,sys_setuid,sys_setgid,sys_setfsuid,sys_setfsgid,sys_set_stbac_attr, sys_kill |
| | $PR_{avai}$ | Sys_setrlimit,sock_recvmsg,sock_sendmsg,sys_brk, schedule, ext3_alloc_block, ext2_alloc_block |

The decision part is a new kernel module that is built for handling requests from the enforcement part. While making a decision, it firstly reads the STBAC flags of subject and object from the information part, and then calls corresponding module rules for deciding whether to permit the access and whether to modify the STBAC flags in the kernel data structure. In the case of denying the access, the decision part will try to redirect the access to the partial copy. If the access is from sys_socket, it will search Trustable-Communication-List to affirm whether the access opens a Trustable-Communication. The process will not be labeled with $F_t$ if the communication is trustable.

The information part saves and maintains all kinds of STBAC flags of subjects and objects. There can be two implementation options: one is to build independently STBAC data structures for saving flags, and the other is to use the existing kernel data structures for saving flags. The main advantage of the former one is that it is independent of Linux kernel codes, but the disadvantage is that it will lose performance significantly; the latter one can use kernel functions to organize and maintain data structures so that it is easy to be implemented and has little performance reduction. The latter one is adopted.

In addition, we created four commands: stbac_set_flag, stbac_get_flag, stbac_admin_trusted_comm and stbac_admin_partial_copy. The stbac_set_flag and stbac_get_flag are used to set and get all kinds of STBAC flags. They can operate on all files and directories under a directory at a time, or operate on all descendants of a process at a time. We also created a shell script named "stbac_init" to automatically initiate and check the STBAC flags for system directories and files when booting the system.

All partial copies are saved under "./stbac". The password and user management commands are modified to synchronize Shared-$O_{inte}$ with its partial copy automatically.

## 4. Model Evaluation

In order to evaluate the STBAC model, we tested the STBAC prototype system from three aspects: protection capability, compatibility and performance. We prepared two Linux machines using RedHat 9.0 whose kernel was 2.4.20-8. One is for attacking, named attacker, with IP 192.1.1.2; the other is the attacked machine, named victim, with IP 192.1.1.1. On the victim, the directories and files to be protected confidentially are /home/szy/data, /etc/passwd and /etc/shadow, while the directories and files to be protected integrally are /boot, /bin, /sbin, /usr/bin, /usr/sbin, /lib, /usr/lib, /dev/kmem and /etc.

### 4.1 Protection Test

Three tests were designed, "remote-user", "web-downloaded-program" and "remote-attack", to verify if STBAC can forbid the remote users who logged in by Non-Trustable-Communications, web-downloaded programs and intruders to illegally access vital resources.

For the convenience of analyzing the test result, the printk() function is called in STBAC-enforced kernel to log every step of intrusion. Function printk() is located in the STBAC decision part, and has the calling form of printk("<4>subject, object, operation, parameter, result"). After every test, we analyzed the logs and drew the dependency graph with the same notations as that in Figure 1.

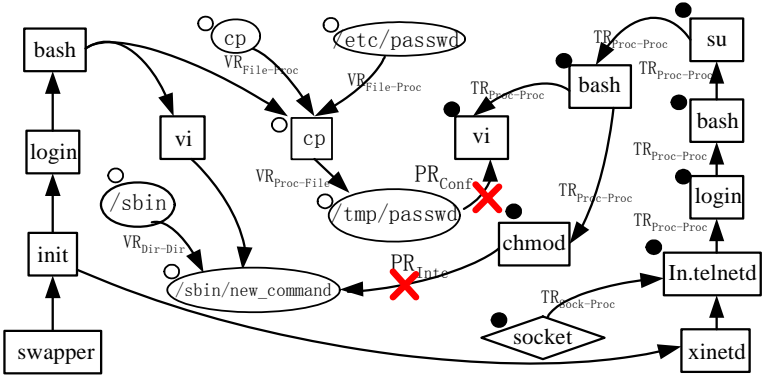
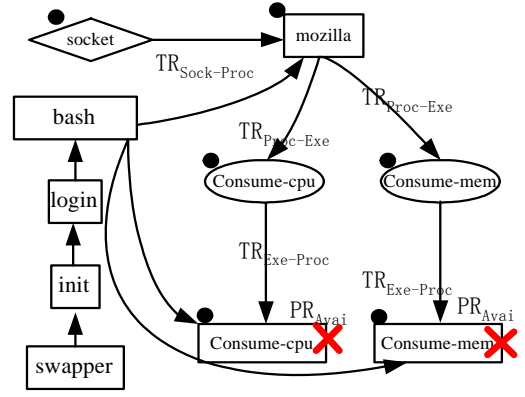

Fig.4. Remote-user test    Fig.5. Web-downloaded program test

*4.1.1 remote-user*

First, a user who logged in as root identity from local created a new file of new_command under /sbin directory, and copied the passwd file to /tmp directory. Then, a user logged in from remote, changed identity to root by su command, tried to modify the new_command file's access right bits, and tried to read the /tmp/passwd file. The test result is shown in Figure 4. The vi process launched by local user's bash is a health subject. It creates a new file /sbin/new_command which inherits $F_{inte}$ flag from parent directory /sbin by $VR_{dir\text{-}dir}$ rule. The cp process is also launched by local user's bash, inherits $F_{leak}$ flag from the cp file when executing the cp file by $VR_{file\text{-}proc}$ rule It inherits $F_{conf}$ flag from /etc/passwd when reading /etc/passwd by $VR_{file\text{-}proc}$ rule, and spreads $F_{conf}$ flag to /tmp/passwd when creating /tmp/passwd by $VR_{proc\text{-}file}$ rule. The remote user's vi process inherits $F_t$ flag by $TR_{sock\text{-}proc}$ and $TR_{proc\text{-}proc}$ rules, then it is refused by $PR_{conf}$ rule when it tries to read /tmp/passwd file. The remote user's chmod process also inherits $F_t$ flag by $TR_{sock\text{-}proc}$ and $TR_{proc\text{-}proc}$ rules, it is refused by $PR_{inte}$ rule when it tries to modify the attribute of the /sbin/new_command file.

So, vital rules can automatically spread vital flags to vital objects and subjects; taint rules can automatically trace remote user's activities in kernel; and protection rules of $PR_{conf}$ and $PR_{inte}$ can forbid remote users to get secrecy or to change integrity information. That is, even if a remote user has gotten root identity through some ways, i.e., Non-Trustable-Communications, his/her illegal activities are still prevented by protection rules.

*4.1.2 web-downloaded-program*

We designed two little programs for downloading. One program named consume-cpu consumes cpu time by an infinite loop; the other program named consume-mem uses up memory by non-stop memory allocation. These two programs were put on the attacker machine and could be downloaded from the web. Figure 5 shows the test result.

As Mozilla process is a $S_t$, consume-cpu and consume-mem downloaded by a local user are both $O_t$ by $TR_{proc\text{-}exe}$ rule. When a user runs consume-cpu program, the new process of consume-cpu becomes $S_t$ by $TR_{exe\text{-}proc}$ rule. When the time exceeds, the process will be stopped by $PR_{avai}$ rule. When the user runs consume-mem, the new process of consume-mem becomes $S_t$ by $TR_{exe\text{-}proc}$ rule. After excessive memory has been allocated, subsequent allocation operations of consume-mem are refused by $PR_{avai}$ rule.

So, taint rules can trace activities of web-downloaded programs in kernel, and $PR_{avai}$ rule can prevent web-downloaded programs from occupying excessive resources.

*4.1.3 remote-attack*

This test uses real attack tools to attack the victim machine from the attacker machine. It takes seven steps: 1) Attack samba to get root shell; 2) Kill syslogd process to stop system log service; 3) Get user's important files to gain user's secrecy; 4) Get passwd and shadow files to crack user's password; 5) Build SUID shell file for convenience of next login; 6) Download and install the hack tool, knark, to leave back-door in the victim; 7) Clean logs.

Samba's version is 2.2.8 and it suffers buffer overflow vulnerability by which the intruder can get root shell. Knark is a famous root-kit whose idea derives from [25], and it will add a kernel module to Linux. Att-samba is an experimental attack tool that is modified from a network-downloaded program. /home/szy/data is an important data file that the intruder wants to get.

The attack progress is shown in Figure 6. Corresponding to the attack steps of 2~6, we can analyze the test result as follows: 2) By $PR_{inte}$ rule, STBAC forbids the tainted process of kill to call function sys_kill() to cease the health process syslogd; 3) By $PR_{conf}$ rule, STBAC forbids the tainted process of cat to read the $O_{conf}$ file /home/szy/data; 4) By $PR_{conf}$ rule, STBAC forbids the tainted process of cat to read the $O_{conf}$ file passwd and shadow; 5) By $PR_{inte}$ rule, STBAC forbids the tainted process of cp to create a new file under /lib which is an $O_{inte}$. Although the intruder creates successfully a SETUID shell under current directory afterwards, the shell file created is an $O_t$. The shell will fail in executing the privileged operation setuid by $PR_{inte}$ rule. 6) By $PR_{inte}$ rule, STBAC forbids the tainted process of mv to create a new file knark.o under directory of /lib/modules/2.4.20-8/kernel/drivers/net which is an $O_{inte}$. At the same time, STBAC forbids the tainted process of insmod to call sys_create_module() function.

As is shown above, attack steps of 2 to 6 all failed. Since these five steps are critical to the intrusion, we can say that STBAC has defeated the intrusion. During this, STBAC uses $TR_{sock-proc}$, $TR_{proc-proc}$ and $TR_{proc-exe}$ to trace activities of the intruder in kernel, and uses $PR_{conf}$ and $PR_{inte}$ to prevent the intruder who has gotten root identity to gain secrecy, to modify files and to leave backdoors.

These three tests validated all of the taint rules, vital rules and protection rules. By taint rules STBAC can trace activities of remote users, web-downloaded programs and intruders in the OS kernel. By vital rules STBAC can spread vital flags to subjects and objects which need protection. Based on spreading taint flags and vital flags, protection rules can prevent invalid actions of $S_t$, thus protecting confidentiality, integrity and availability.

### 4.2 Compatibility Test

On the STBAC-enforced Linux kernel 2.4.20-8, we have run many network applications and local applications without incompatible problems. Table 3 shows the applications. For testing remote management, we used Webmin and SSH to manage user accounts, file system, network and services from remote computers. For testing remote upgrading, we used up2date to upgrade RedHat Linux from RedHat website. As Webmin, SSH and up2date all have authentication and encrypted communication, i.e., they run on Trustable-Communications, STBAC does not restrict any action of them.

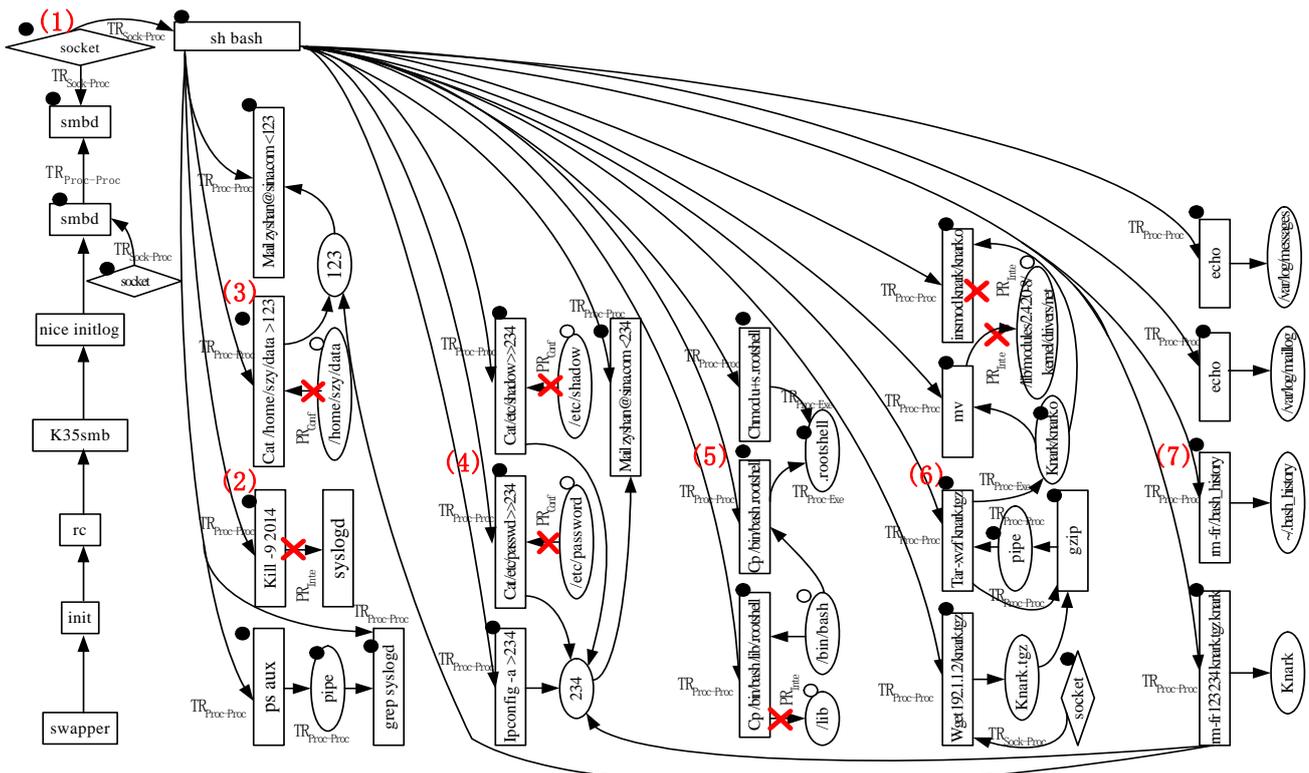

**Fig.6.** Remote-attack test

**Table 3** Compatibility test

| Network application | | | Local application | | |
|---|---|---|---|---|---|
| Network services | Remote management | Remote Upgrading | Development | Desktop | System management |
| apache, mysql, samba, ftp, telnet, mozilla, dhcp, rlogin, sendmail | Webmin (use ssl), ssh | up2date (use ssl) | Kernel compiling, gcc, gdb, vi, QT | KDE, GNOME | df, top, free, useradd, pstree, mount, tar, passwd, etc. |

### 4.3 Performance Test

STBAC has little impact on Linux performance. First, the decision rules are simple as they only compare flags on the subject and object; second, getting decision data is also fast in that decision data are saved in the control block of the current process or inodes of the currently opened files or directories. In the performance test, we compared the performance of two kernels: the original Linux kernel and the STBAC-enforced kernel. The comparison was done in two environments: non-$S_t$ environment and $S_t$-existing environment.

*4.3.1 Non-$S_t$ Environment*

In this test, we applied the "kernel compile" [29] testing approach. The "kernel compile" is a broadly accepted method for testing the general performance of Linux. The test uses "victim" as the test machine, and uses the default configuration for kernel compiling. The victim is a 1.1GHz Intel Celeron machine with 256M sized main memory, and 100MHz DRAM clock.

**Table 4** Test results for compiling Linux kernel (Sec)

| Kernel | Time categories | 1 | 2 | 3 | Average |
|---|---|---|---|---|---|
| Non-STBAC | Real | 444.652 | 426.104 | 425.167 | 431.9743 |
| | User | 393.370 | 393.780 | 392.680 | 393.2767 |
| | System | 27.550 | 27.280 | 28.470 | 27.76667 |
| STBAC | Real | 445.804 | 427.457 | 427.894 | 433.7183 |
| | User | 393.890 | 393.680 | 394.070 | 393.88 |
| | System | 28.340 | 28.100 | 28.280 | 28.24 |

Table 4 shows the results. Non-STBAC and STBAC indicate respectively the original Linux kernel and the STBAC-enforced Linux kernel. Unlike [29] and [28], we recorded the User-time and System-time as well as the Real-time. As the model is implemented in kernel and the Real-time is easily influenced by random environment factors, we focus on the System-time instead of the Real-time. In Table 4, the System-time of the STBAC-enforced kernel increased 1.7% compared with that of the non-STBAC kernel. The comparison test was run three times. Similar results were obtained. The first run of the test was done with "cold-cache" while the other two were in "warm-cache". So the Real-time in the first run was more than that of the other two, since the first compiling has to read a great deal of files from disk while the last two compilings can read many files from memory or disk swapping area.

*4.3.2 $S_t$-existing Environment*

As $S_t$ comes from network communication, we designed two UDP communication programs for this test. One was installed on the victim machine as a client; the other was installed on the attacker machine as a server. The client sent UDP request package to the server. After receiving the answer package from the server, the client resent the same UDP package to the server. It repeated sending 100,000 packages to the server and receiving 100,000 packages from the server.

The result showed that each cycle consumed 124.4 microseconds on the non-STBAC kernel and 130.1 microseconds on the STBAC-enforced kernel. Compared with the non-STBAC kernel, the STBAC-enforced kernel demonstrated a 4.6% increase of the consumed time.

Based on the above tests, we can safely say that the STBAC-enforced Linux can protect effectively important directories, files and processes without significant impact on compatibility and performance. Performance reduction is only around 1.7% to 4.6%.

## 5. Related Works
### 5.1 Relations with DTE

DTE, proposed by Lee Badger et al. [30][1], is implemented in Linux by Serge Hallyn et al. [28], and is also adopted by SE-Linux [22]. It groups processes into domains, and groups files into types, and restricts access from domains to types as well as from domains to other domains. In a predefined condition, a process can switch its domain from one to another.

STBAC can be viewed as a type of dynamic DTE. It divides all processes into two domains: $S_t$ and $S_h$, and divides all objects into five types: $O_t$, $O_h$, $O_{inte}$, $O_{conf}$ and $O_{avai}$; It defines access rights of each domain: $S_h$ can access any object; $S_t$ can access any object except reading $O_{conf}$, writing $O_{inte}$ and allocating excessive $O_{avai}$; $S_h$ can switch to $S_t$ by taint rules.

Dynamic characteristic of STBAC is reflected in that both domains and types in STBAC can change dynamically during the system execution, but in DTE only domains can change and types cannot change. Domains and types change in STBAC according to the taint rules, which are automatically triggered by the intruder's activities in the system. However, domain changing in DTE takes place when executing the entry point file that needs administrator's predefinition.

Due to this dynamic characteristic of STBAC, administration work is dramatically decreased. Users do not need to predefine which subject is $S_h$ or $S_t$, and which object is $O_h$ or $O_t$. These definitions are automatically done by the taint rules during system execution.

### 5.2 Intrusion Backtracking in OS

Another related work is the intrusion backtracking in OS. In 2003, S.T.King and P.M.Chen built an effective intrusion backtracking and analyzing tool named Backtracker [13][10]. It can help administrators to find intrusion steps with the help of the dependency graph that is generated by logging and analyzing OS events. Zhu and Chiueh built a repairable file service named RFS [11], which supports kernel logging to allow roll-back of any file update operation, and keeps track of inter-process dependencies to quickly determine the extent of system damage after an attack/error. In 2005, Ashvin Goel and Kenneth Po built an intrusion recovery system named taser [12], which helps in selectively recovering file-system data after an attack. It determines the set of tainted file-system objects by creating dependencies among sockets, processes and files based on the entries in the system audit log. These works all focus on the intrusion analysis and recovery by logging system activities, and directly inspired the taint rules of STBAC. The most distinctive point in our work is that our objective is to build an access control mechanism that can trace and block intrusions in real time.

### 5.3 Tainted Data Analysis

The third related work is tainted data analysis. In 2005, James Newsome and Dawn Song built a detection tool named TaintCheck [26], which performs dynamic taint analysis on a program by running the program in its own emulation environment. In 2006, Alex Ho and Michael Fetterman et al. built a taint-based protection system [27] that traces tainted data in a Virtual-Machine-and-hardware-emulation-combined environment. STBAC differs from these two works in the following ways: (1) the meaning of the taint is different. Taint in STBAC isn't tainted data, but tainted OS objects or subjects, such as tainted processes or tainted files; (2) the method of tracing taint is different. STBAC traces taint on the basis of the operations between OS-level subjects and objects, for example processes reading files or sockets. But these two works are both based on hardware-level instructions, such as LOAD, STORE and MOVE; (3) the objective is different. STBAC tries to prevent illegal actions of tainted subjects, while these two works try to detect intrusions.

## 5.4 Information-Flow-Based Access Control

Asbestos [31] and HiStar [20] are both information-flow-based access control. They label data with its owner, track information as it moves around in OS, and base access control decision on the labels. Our work differs from Asbestos' in two points. First, Asbestos aims to isolate users from each other, thus to prevent illegal access to user data; STBAC exploits DAC to isolate users and uses mandatory protection rules to forbid $S_t$ to illegally access vital resources which contain not only user data but also important system files, directories and privileges. Second, Asbestos uses user-related labels to trace data flow. This has to use "event process" and "label", which increase complexity and consume excessive resources. STBAC uses simple flags and taint rules to trace $S_t$'s activities, so the implementation is simple and impact to performance is kept at a minimum.

## 5.5 Linux Security Enhancement

There are several famous Linux security enhancement projects, such as SELinux [22], LIDS [4], DTE [28], systrace [21], LOMAC [2][3], and etc. SELinux is a powerful Linux security enhancement project. It can flexibly support multiple security policies. But for general users, it is difficult to bring into play, because it requires professional knowledge on the part of the user. LOMAC has similar ideas with ours. It implements the Low-Water-Mark model [6] in Linux kernel, and aims to bring simple but useful MAC integrity protection to Linux. It maintains good compatibility with existing software. But LOMAC does not consider safeguarding confidentiality and usability.

If having interest, please also read our other papers [32]-[71].

## 6. Conclusions

In this paper, we present a new OS access control model named STBAC. It consists of four parts: Taint, Health, Vital and Protection. The Taint might be controlled by an intruder and consists of $S_t$, $O_t$ and taint rules which can trace activities of $S_t$. The Vital should be protected properly as it represents vital resources which are the valuable user data and the foundation for system to provide services, hence vital resources usually become the final targets of an intrusion. The Protection consists of three mandatory protection rules that forbid $S_t$ to illegally access vital resources. The Health is not tainted nor labeled as vital ones, and it can access the Taint and the Vital.

STBAC have reached its four goals: protecting vital resources, compatible with existing software, simple to administer and good performance. For protecting vital resources, analysis shows that STBAC is capable of forcibly protecting confidentiality, integrity and partial availability. The three protection tests further validated this experimentally. For compatibility, analysis shows that STBAC does not influence the actions of local users and remote users using Trustable-Communications. Remote administrator can still manage the computer and upgrade application software through Trustable-Communications. In addition, STBAC does not influence most actions of $S_t$, because it only forbids $S_t$ to illegally access vital resources. Compatibility exceptions come from Shared-$O_{conf}$ and Shared-$O_{inte}$, which are of tiny amounts and Trustable-Communication-List and Partial-Copy mechanisms can be used to resolve them. The test on application compatibility validated this goal experimentally. For simplicity, analysis shows that the main administration work of STBAC is to set vital flags for user files and directories that need to be protected, which is straightforward and easy to understand. The vital flags of system files and directories can be automatically labeled by a shell script "stbac_init" when booting the system. $F_t$ is automatically generated and propagated by kernel and does not require any manual operation. For performance, tests in both non-$S_t$ environment and $S_t$-existing environment showed that there is merely 1.7%~4.6% performance reduction caused by STBAC.

Therefore, the STBAC model is useful in OS to defeat network attacks while maintaining good compatibility, simplicity and system performance.